\documentclass[twocolumn,showpacs,epsfig]{revtex4}
\usepackage{graphicx}

\begin{document}

\draft
\title{Survival probability time distribution in dielectric cavities}
\author{Jung-Wan Ryu$^{1,2}$}
\author{Soo-Young Lee$^1$}
\author{Chil-Min Kim$^1$}
\author{Young-Jai Park$^2$}
\affiliation{$^1$ National Creative Research Initiative Center for Controlling Optical Chaos,\\
Pai-Chai University, Daejeon 302-735, Korea}
\affiliation{$^2$ Department of Physics, Sogang University, Seoul 121-742, Korea}

\begin{abstract}
We study the survival probability time distribution (SPTD) in dielectric cavities.
In a circular dielectric cavity the SPTD has an algebraic long time behavior, $\sim t^{-2}$
in both TM and TE cases, but shows different short time behaviors due to the existence of 
the Brewster angle in TE case where the short time behavior is exponential.
 The SPTD for a stadium-shaped cavity decays exponentially, and
the exponent shows a relation of $\gamma \sim n^{-2}$, $n$ is the refractive index,
 and the proportional coefficient 
is obtained from a simple model of the steady probability distribution. 
 We also discuss about the SPTD for a quadrupolar deformed cavity and
show that the long time behavior can be algebraic or exponential depending on
 the location of islands.
\end{abstract}
\maketitle
\narrowtext

\section{Introduction}

Recently, lasing modes from dielectric microcavities have attracted much attention
due to its potential application to optoelectric circuits and optical communications \cite{Cha96}.
In particular, there was a lot of theoretical and experimental effort 
to excite directional lasing modes in deformed microcavities \cite{Noc97}.
It is now well known that the lasing pattern has a very close relationship with
the internal ray dynamics given by the boundary geometry of cavity.
It is also reported that the property of the openness of the dielectric cavity 
plays an important role in the resonance pattern analysis \cite{Lee04,Har04}.

For a general open system, the survival probability time distribution (SPTD)
or its time derivative, the escape time distribution, is a basic physical quantity. 
Many studies are focused on the relation between the long time behavior of the SPTD and
the internal dynamics, and it is known that the SPTD has algebraic and
exponential decays in integrable and chaotic systems, respectively
\cite{Bau90,Dor90,Fri01,Bun05,Alt96,Sch02}.
In mixed systems, having both integrable islands and chaotic sea in phase space,
the SPTD has algebraic long time behavior which originates from the slow
escape mechanism due to the stickiness of KAM tori \cite{Fen94}.

The property of openness of the dielectric cavity is different from the
open systems previously studied \cite{Bau90,Dor90,Fri01,Bun05,Alt96,Sch02}.
Rays can escape through any boundary
point, and  partial escapes, depending on the incident angle, are possible.
This unique property can be reflected on the long time behavior of
the SPTD.

In this paper, we study the SPTD  in dielectric cavities of various
boundary geometries such as circle, stadium, and quadrupole, which are
typical examples of integrable, chaotic, and mixed systems, respectively.
The SPTDs in these dielectric cavities show basically similar behavior to
the open cavity with a small hole on the cavity boundary, but the exponents
are different. In particular, we show that the ergodic property cannot be
applied for the stadium-shaped dielectric cavity even in the small opening
limit, $n\rightarrow \infty$, $n$ is the refractive index. 

The paper is organized as follows. In Sec. II the algebraic long time behavior of
the SPTD in the circular dielectric cavity is derived analytically  and confirmed
numerically for both TM and TE waves. 
It is shown in Sec. III that the SPTD for a stadium-shaped cavity decays exponentially, 
and the exponent $\gamma$  has $\sim n^{-2}$ dependence and the proportional coefficient  
can be understood from a simple model of the steady probability distribution (SPD).
The SPTD in the quadrupole-deformed dielectric cavity is discussed in Sec. IV and
we finally summarize the results in Sec. V.

\section{Circular dielectric cavity - integrable system}

Many authors have studied the SPTD
for the open billiard with a small hole on boundary \cite{Bau90,Dor90,Fri01,Bun05}.
It is known that the SPTD in a circular billiard decays algebraically,
$P_{sv}(t) \propto t^{-1}$.
In this section we study the SPTD for the circular dielectric cavity,
and it will be shown that the SPTD shows still an algebraic decay but the exponent
is different.

For simplicity, we focus on TM  wave first and for TE wave we will mention
only difference later. In the circular geometry, ray dynamics is integrable and
rays in the open area of the phase space, i.e., $-1/n < p < 1/n$, $p=\sin \theta$,
 $\theta$ being the incident angle,  can partially 
escape from the cavity. 
The formal expression of the SPTD in the circular dielectric cavity is given by
\begin{eqnarray}
P_{sv}(t) =\frac{n}{2L} \int_{0}^{L}ds  \int_{-p_c}^{p_c}dp \,\, R(p)^{N(t)},
\label{psv}
\end{eqnarray}
where $L$ is the boundary length, and $p_c$ is the critical line 
for total internal reflection, i.e., 
$p_c=\sin \theta_c=1/n$, and  $R(p)$ is the reflection coefficient for TM wave \cite{Haw95},

\begin{equation}
R(\theta)=\left({{n \cos{\theta} - \cos{\theta_t}}
\over{{n\cos{\theta} + \cos{\theta_t}}}}\right)^2,
\label{Rtheta}
\end{equation}
where $n \sin \theta = \sin \theta_t$, and
$N(t)$ is the number of bounce on the boundary. Since $N(t)=t/2\cos \theta$ in the
circular geometry, when considering a unit circle and
a time scale as the length of ray trajectory, Eq.(\ref{psv}) can
be rewritten as   
\begin{eqnarray}
P_{sv}(t) = n\int_{0}^{\theta_c} d\theta \,\, \cos{\theta}  \exp\left[{-G(\theta)t}\right],
\label{pG}
\end{eqnarray}
where
\begin{equation}
G(\theta)\equiv {{1} \over {\cos{\theta}}}
 \ln \left({1+ {{2\cos{\theta_t}} \over {n\cos{\theta} - \cos{\theta_t}}}}\right).
\end{equation}

Note that the rays near the critical line $p_c$ can survive longer time and
dominate long time tail behavior. Therefore, we can expand $G(\theta)$ from
$\theta_c$ by changing variable, $\theta=\theta_c - \chi$, as
\begin{eqnarray}
G(\theta) \approx \alpha \chi^{1/2} + \beta \chi^{3/2}+ \cdots ,
\label{G}
\end{eqnarray}
where
\begin{equation}
\alpha = {{2n\sqrt{2\sqrt{n^2 - 1}}}\over{n^2 - 1}}
\end{equation}
and
\begin{equation}
\beta =-\alpha (\frac{n^2-6}{4\sqrt{n^2-1}}-\frac{2}{n}). 
\end{equation}
Substituting the lowest term in Eq.(\ref{G}) into Eq.(\ref{pG}), we can obtain the long time
behavior of the SPTD as
\begin{eqnarray}
P_{sv}(t) &\simeq& 
  \frac{2\sqrt{n^2-1}}{\alpha^2} t^{-2}  \left[ 1-(1+ \alpha \sqrt{\theta_c}t)
e^{-\alpha \sqrt{\theta_c}t} \right] \nonumber \\
 &\simeq& \frac{2\sqrt{n^2-1}}{\alpha^2} \, t^{-2}.
\label{t2}
\end{eqnarray}
We emphasize that  the SPTD for the circular dielectric cavity decays as $t^{-2}$
as shown in Eq.(\ref{t2}), different from 
the open billiard with a small hole where decays as $t^{-1}$. This means
that the property of openness can change the exponent of
the algebraic decaying SPTD.

For TE wave case the reflection coefficient  is given by
\begin{equation}
R_{TE}(\theta)=\left({{n \cos{\theta_t} - \cos{\theta}}\over{{n\cos{\theta_t} + \cos{\theta}}}}\right)^2,
\label{RTE}
\end{equation}
and the expansion of $G(\theta)$ and the SPTD at a long time are
the same as Eq.(\ref{G}) and Eq.(\ref{t2}) with different
expansion coefficients, i.e.,
\begin{equation}
\alpha_{TE}={{2n^3 \sqrt{2 \sqrt{n^2 - 1}}}\over{n^2 - 1}}=n^2 \alpha
\end{equation}
and
\begin{equation}
\beta_{TE}=-\frac{\alpha_{TE}}{4\sqrt{n^2-1}} (8n^4 + n^2 + 6).
\end{equation}
We note that the dependence of the SPTD on the refractive index $n$ in TM and TE waves is
quite different, i.e., $P_{sv}(t)\simeq n^2 t^{-2}$ for
TM case, but $P_{sv}(t)\simeq n^{-2} t^{-2}$ for TE case.
The proportionality of  $n^{-2}$ of the TE case  does not mean that
the circular cavity with a higher $n$ is more leaky, since we take into
account of only the open region in the phase space,$-1/n <p<1/n$ (see Eq.(\ref{psv})).

\begin{figure}
\begin{center}
\includegraphics[width=0.4\textwidth]{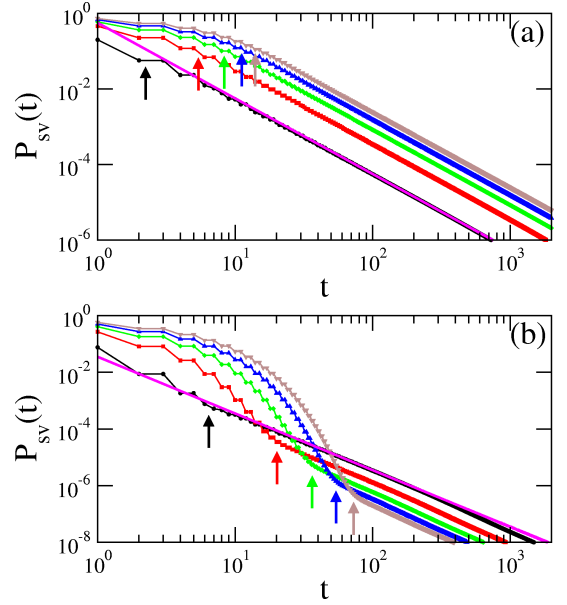}
\caption{(color online) The SPTDs of the circular dielectric cavity 
for (a) TM and (b) TE waves.
Black circle, red square, green diamond, blue triangle (up), and brown triangle (down) are
for n=2, 4, 6, 8, and 10, respectively.
These shows $t^{-2}$ long time behavior and a very good agreement with the solid lines
which represent the $n=2$ case in Eq.(\ref{t2}). 
}
\end{center}
\end{figure}

In order to perform numerical calculation for the SPTD in the circular cavity,
 we take $10^8$ random initial
points in the open region of the phase space. We then trace each point
with a weight determined by $R(p)$ when bouncing from the boundary, 
and sum the weights between $t$ and $t+\Delta$, we take $\Delta=1$ in the calculations,
for all points in the ensemble, and finally normalize to be unit when $t=0$.
Figure 1 shows the numerical results of the SPTD in the circular cavity for TM and TE cases.
It is clear that  the SPTD for both cases shows
an algebraic long time behavior, $\sim t^{-2}$, and the dependence on $n$ 
is correctly described by Eq.(\ref{t2}) which is indicated by the solid lines for $n=2$
in Fig.1.

\begin{figure}
\begin{center}
\includegraphics[width=0.4\textwidth]{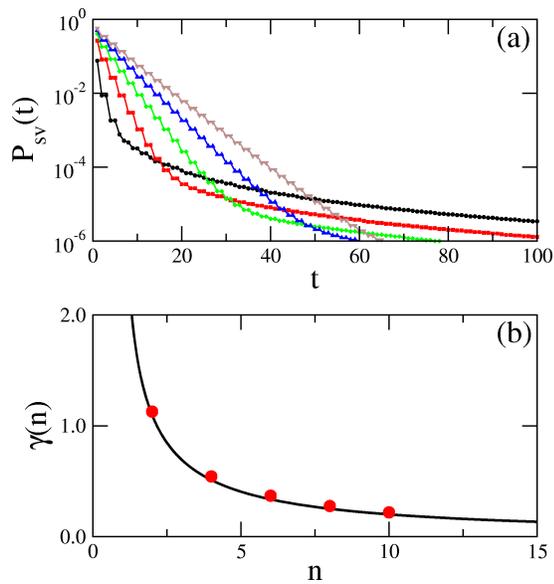}
\caption{(color online) (a) The exponential short time behavior of the SPTDs for TE case.
The colors for different $n$ is the same as in Fig. 1. These can be fitted as
 $\exp(-\gamma(n) t)$ in a short time range.
(b) The exponents $\gamma(n)$. The solid line represents the result of a simple 
approximation of Eq.(\ref{gamma}) .}
\end{center}
\end{figure}

 A substantial difference between the TM and TE cases appears in the short time
behavior. As shown in Fig. 1 (a) the short time behavior of the TM SPTD is smoothly
connected to the $ t^{-2}$ long time tail, on the other hand that of the TE SPTD
shows rather an abrupt transition to the algebraic long time tail and the detail of
the short time behavior seems to be characterized by an exponential decay.
This exponential short time decay is clear in Fig. 2 (a) and the exponent
$\gamma(n)$, when fitted as $\exp(-\gamma(n) t)$, appears as the solid dots in Fig. 2 (b).
These exponents are well described by a simple approximation for the reflection
coefficient $R_{TE} (\theta)$. If we expand  $R_{TE} (\theta)$ at $\theta=0$ and take
only the lowest term, then
\begin{equation} 
P_{sv}(t)\simeq \exp(-\ln(\frac{n+1}{n-1}) t) =\exp(-\gamma(n) t).
\label{expand0}
\end{equation}
The solid line in Fig. 2 (b) represents the relation
\begin{equation}
\gamma(n)=\ln(\frac{n+1}{n-1}).
\label{gamma}
\end{equation} 
Here we emphasize that although the lowest term of the expansion of Eq.(\ref{expand0})
is the same for both TM and TE cases, only TE case allows the exponential short time
decay. The reason for this is the existence of the Brewster angle in the TE case, 
$\theta_B=\arctan(1/n)$ where rays can escape without reflection, i.e., $R(\theta_B)=0$. 
The rays with the incident angle in the range of $-\theta_B < \theta < \theta_B$
dominate the short time behavior, while the other parts, $\theta_B < |\theta| < \theta_c$,
mainly contribute to the long time algebraic tail.

It is important to know when the algebraic decay starts to appear in both TM and TE cases.
In TM case, the main factor for the deviation from the $t^{-2}$ decay comes from the
finite integral bound, and it corresponds to the terms containing the upper bound
 $\theta_c$ in Eq.(\ref{t2}). We then estimate the transition time when the deviation from
the $t^{-2}$ decay is about 10 \%, and the result is
\begin{equation}
t_c \simeq \frac{1.38 (n^2-1)^{3/4}}{n\sqrt{\arcsin(1/n)}} \propto n.
\end{equation}
In Fig. 1 (a) the corresponding transition times are indicated by arrows and
show a good agreement with the numerical calculations for various $n$.

Due to the existence of the Brewster angle, the transition time for TE SPTD
can be determined by a different way. As mentioned above, the TE SPTD shows
a short time exponential behavior and a long time algebraic behavior.
Therefore we can estimate the transition time by finding the intersection time
for both different behaviors.
From Eq.(\ref{t2}) with $\alpha_{TE}$ and Eq.(\ref{expand0}),
 for a large $n$ we can get an implicit equation for the transition time
as
\begin{equation}
\frac{t_c}{n}\exp(-\frac{t_c}{n})= \frac{1}{2n^2}.
\label{tcte}
\end{equation} 
The transition times for various refractive indices are indicated by arrows in Fig. 1 (b)
and well represent the transition
times of the numerical results. The solution $t_c$ of the above equation
cannot be described by a simple power of $n$, but we can show 
\begin{equation}
t_c (n) \propto n^{\mu(n)}, \,\,\,\, \mu(n) > 1.
\end{equation}
If we take a logarithm of Eq.(\ref{tcte}), then we get
\begin{equation}
\frac{t_c}{n}-\ln \frac{t_c}{n} = \ln 2 n^2,
\end{equation}
which generally has two solutions and the larger solution is relevant.
The point $t_0$, at which the slopes of the two functions in the left hand side of
the above equation are identical, should locate between the two solutions.
By differentiating the above equation, we get $t_0=n$.
Therefore,
\begin{equation}
t_c > t_0 = n.
\end{equation}

Even though both TM and TE cases show the same $t^{-2}$ long time
decay in the circular cavity,
 the short time behavior and the $n$ dependence of the transition time
are quite different. We emphasize that these differences originate
from the existence of the Brewster angle in the TE case.

\section{Stadium-shaped dielectric cavity - chaotic system}

As an example of chaotic dielectric cavities, we take a stadium-shaped one 
with parallel linear segments of a length $l$ and two semicircles of a radius $R$.
The stadium-shaped billiard has been a typical chaotic system in the research of   
classical and quantum chaos. The escape property through a small hole on the
boundary of the stadium-shaped billiard has been investigated by many authors \cite{Alt96}.  
They have shown that the escape time distribution exponentially decays first and later
becomes algebraic, and the transition time $t_c$ increases as the hole size
decreases. The algebraic decay at long times comes from the stickiness near the marginally
stable line in phase space corresponding to the bouncing ball trajectories.
On the other hand, in the stadium-shaped dielectric cavity the ray trajectories
of the bouncing ball type cannot contribute to the long time behavior due to
the property of openness, i.e., rays with almost vertical incidence escape
easily and contribute to the short time behavior. As a result the SPTD shows only
exponential decay (see Fig. 4, 5).

The exponential decay in the dielectric chaotic cavity implies the existence
of the {\it steady probability distribution} (SPD), $P_s(s,p)$ which is defined as the
spatial part of the survival probability distribution $\tilde{P}_{sv}(s,p,t)$ \cite{Lee04}.
With this SPD, we can express the SPTD as
\begin{equation}
P_{sv}(t)=\int_0^L ds \int_{-1}^{1} dp \,\, \tilde{P}_{sv}(s,p,t)
 \simeq C \exp(-\gamma t),
\end{equation}
where $C$ is a constant and
\begin{equation}
\gamma =\int_0^L ds \int_{-1}^{1} dp \,\, P_s (s,p) T(p),
\label{gammaspd}
\end{equation} 
where the transmission coefficient $T(p)$ is given as $T(p)=1-R(p)$.

\begin{figure}
\begin{center}
\includegraphics[width=0.4\textwidth]{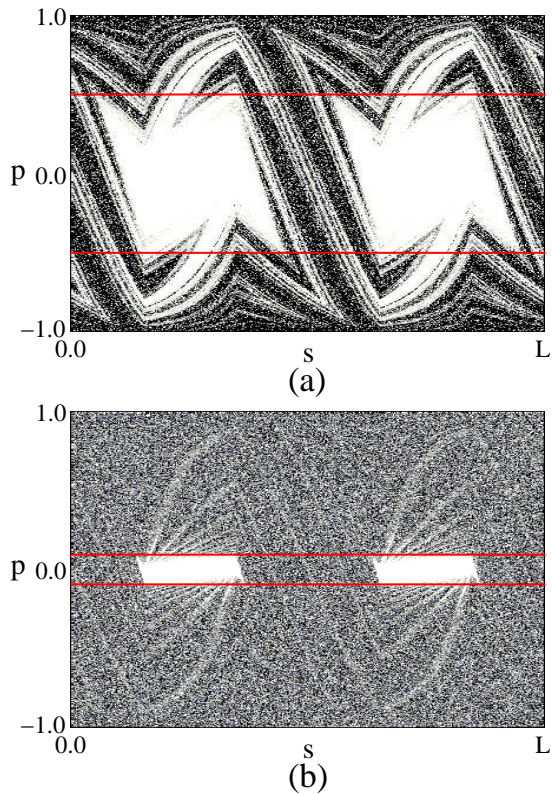}
\caption{The steady probability distributions. (a) $n=2$ case with system parameters $(R,l)=(1,2)$
 for TM wave.  (b) $n=10$ case with system parameters $(R,l)=(1,2)$ for TM wave.
The black points indicate the rays traced from a random ensemble with weights higher than 0.1
at about $t=12$ in $n=2$ case and $t=37$ in $n=10$ case.
The solid lines denote the critical line $\pm p_c$ for total internal reflection.
}
\end{center}
\end{figure}

Note that the above equation is satisfied in the exponential decay region and
cannot describe the nonexponential very short time behavior. 
From Eq.(\ref{gammaspd}), if we know the SPD, we can estimate the decay rate $\gamma$.
However, the structure of the SPD is usually very complicated because it depends
on the openness as well as the boundary geometry of the cavity. 
Figure 3 (a) shows the approximate of the SPD when $n=2$ which is a
snap shot of the $\tilde{P}_{sv}(s,p,t)$ captured at about $t=12$.
The partial escape property of the dielectric cavity  allows for rays
to distribute on unstable manifold structure in the open region, $-1/n < p < 1/n$.

\begin{figure}
\label{sttm}
\begin{center}
\includegraphics[width=0.4\textwidth]{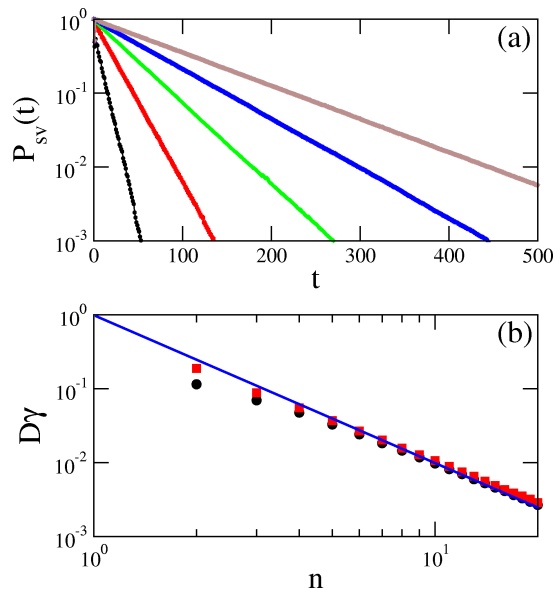}
\caption{ (color online) The SPTD and the decay rates for TM wave.
(a) The exponential SPTD for a stadium dielectric cavity with $(R,l)=(1,1)$.
Black, red, green, blue, and brown lines are for n=2, 4, 6, 8, and 10, respectively.
(b) The decay rates with increasing $n$. Black circle and red square represent
results for system parameters $(R,l)=(1,1)$ and $(1,2)$, respectively.
The solid line shows $n^{-2}$ behavior. The coefficient $D$ is given as
 $D=\frac{\sqrt{A}L}{2\pi^2R}$. 
}
\end{center}
\end{figure}

Even though it is difficult to estimate the SPD in usual cases, for the large $n$ case,
the small opening case, we can simplify the SPD by assuming a uniform distribution
over the whole phase space except the open regions related to the linear segments
of the stadium boundary. The approximate of the SPD for $n=10$ shown in Fig. 3 (b)
supports this assumption. We note that this is a substantial difference from the
escape through a small hole on boundary where entirely uniform distribution is
assumed due to the ergodic property \cite{Bau90}.
 Based on the assumption of the partial ergodicity,
we can rewritten the decay rate as
\begin{equation}
\gamma=\frac{\pi R}{\sqrt{A}(L-2l/n)}\int dp \,\, T(p),
\label{gammaT}
\end{equation}
where we insert the factor $1/\sqrt{A}$, $A=\pi R^2+2Rl$ being the area of the stadium,
from the consideration of time scale. 
The integral of the above equation means the degree of openness and in the large $n$ limit
decreases as $\sim 2\nu \pi n^{-2}$ for both TM ($\nu=1$) and TE ($\nu=2$) cases (see Appendix).
Therefore, for the large $n$ limit the decay rate becomes
\begin{equation}
\gamma \simeq  \frac{2\nu\pi^2 R}{\sqrt{A}L} n^{-2}.
\label{gammafinal}
\end{equation}

Numerical results for the SPTD in the chaotic stadium-shaped dielectric cavity are shown
in Figs. 4 and 5 for TM and TE cases, respectively. We take two systems; one is $(R,l)=(1,1)$
and the other is $(R,l)=(1,2)$. In calculation, we use a random ensemble of $10^4$ initial
points spread over the whole phase space and trace the survival probability with time, the time
is scaled as the length of trajectory in the spatial space as before.
The exponential behavior of the SPTD is clear even at long time limit in both TM and TE cases.
This means that the sticky region locating on the center of the open region in phase space 
dose not contribute long time decay due to its easy escape.  The dependence of the decay rate
$\gamma$ on the refractive index $n$ shows a very good agreement with Eq.(\ref{gammafinal})
for large refractive indices, in both systems with different area $A$.
 This implies that even in the small opening limit, $n \rightarrow \infty$,
we cannot use the ergodic property over the whole phase space. Instead, we have to consider
structure of the SPD even in the small opening limit.

\begin{figure}
\label{stte}
\begin{center}
\includegraphics[width=0.4\textwidth]{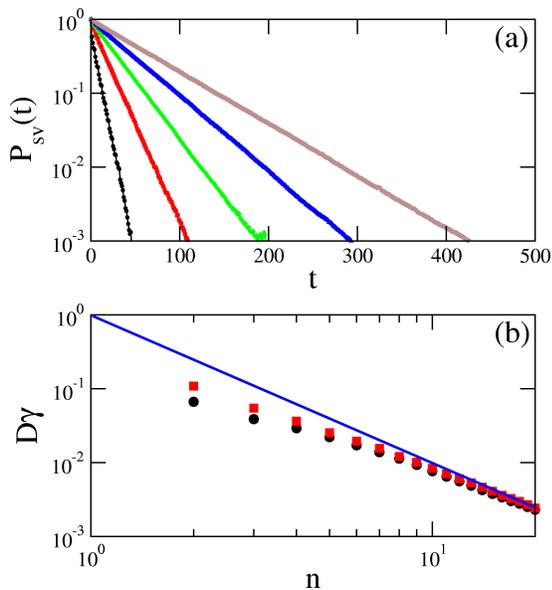}
\caption{(color online) The SPTD and the decay rates for TE wave.
Detail of caption is the same as Fig. 4 except $D=\frac{\sqrt{A}L}{4\pi^2R}$ .
}
\end{center}
\end{figure}

\section{Quadrupole-deformed dielectric cavity - mixed system}

The escape property in generic mixed systems, showing a mixed phase space portrait:
integrable islands in a chaotic sea, has been extensively studied.
It is well known that the long time behavior of the SPTD is
algebraic due to the stickiness of the KAM tori surrounding islands
\cite{Fen94},
\begin{equation}
P_{sv} (t) \sim t^{-\eta}.
\end{equation}
However, there is no rigorous theory expecting the value of the exponent $\eta$
which has been estimated based on numerical calculations and seems to be nonuniversal.
 
In this section, we consider a quadrupolar dielectric cavity which is
the typical example of a deformed microcavity and shows a mixed dynamics. 
The boundary equation is, in the polar coordinates,
\begin{equation}
r(\phi)=1+\varepsilon \, \cos2\phi,
\end{equation}
where $\varepsilon$ is the deformation parameter. 
Here, we present numerical results of the SPTD and show that the long time behavior
of the SPTD is determined by whether islands locate in the closed region, $p_c < |p| < 1$,
or not.

\begin{figure}
\begin{center}
\includegraphics[height=0.35\textwidth,width=0.4\textwidth]{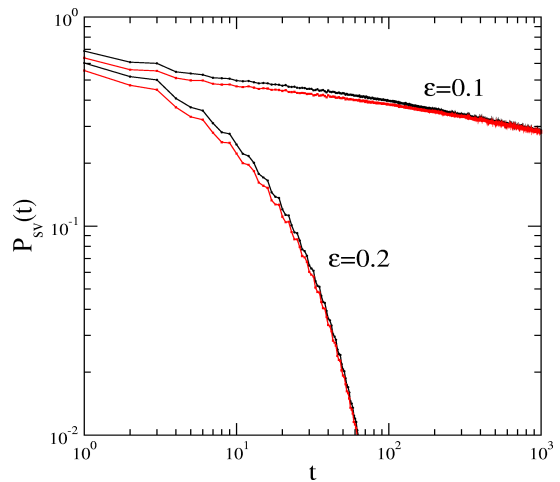}
\caption{(color online) The SPTDs in the quadrupolar deformed dielectric cavity.
The SPTDs are algebraic in the  $\varepsilon=0.1$ cases while exponential in the  $\varepsilon=0.2$ cases.
The black and red lines denote the case of  TM and TE waves, respectively.}
\end{center}
\end{figure}

For $n=2$ case, we numerically calculate the SPTDs at two deformation parameter values, 
$\varepsilon=0.1$ and $0.2$, which are shown in Fig. 6.
In the $\varepsilon=0.1$ case, the SPTD shows an algebraic decay, i.e., 
$P_{sv} (t) \sim t^{-0.2}$, which is consistent with the previous studies on mixed systems.
However, in the $\varepsilon=0.2$ case, the long time behavior of the SPTD is exponential,
i.e., $P_{sv} (t) \sim \exp (-0.05t)$. 
This clear difference of the SPTD between $\varepsilon=0.1$ and $0.2$ cases can be
explained by the phase space portraits.
Figure 7 (a) shows the phase space portrait for $\varepsilon=0.1$ case.
There are many islands in the closed region, $p_c < p < 1$, so the stickiness of 
the KAM tori delays the ray escape and results in the algebraic tail. 
On the other hand, as shown in Fig. 7 (b) there is no island in the closed region
for $\varepsilon=0.2$ case and all islands exist in the open region.
The rays trapped by the stickiness of the KAM tori contribute to the short time 
escape behavior and the resulting SPTD shows exponential long time decay.
Therefore, the position of islands plays important role to understand the SPTD of mixed systems.

\begin{figure}
\begin{center}
\includegraphics[width=0.4\textwidth]{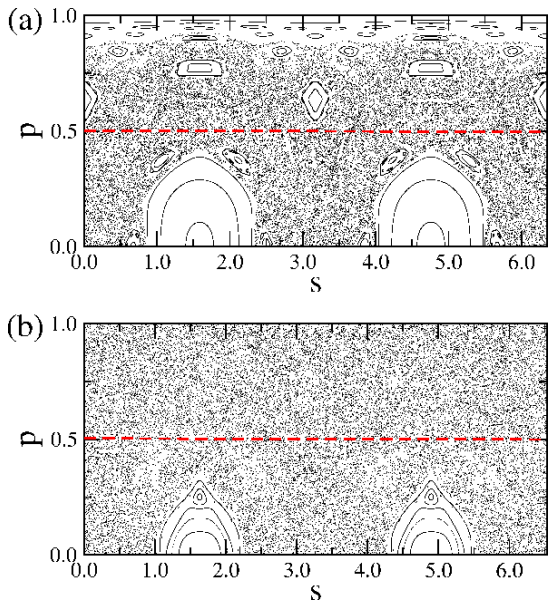}
\caption{ The phase space portraits for the quadrupolar deformed cavity.
(a) The $\varepsilon = 0.1$ case. 
(b) The $\varepsilon = 0.2$ case.
The dashed lines denote the critical line $p_c=1/2$ for total internal reflection
for $n=2$.}
\end{center}
\end{figure}

\section{Summary}

We have investigated the survival probability time distribution (SPTD) in three dielectric cavities
showing different ray dynamics; circle (integrable), stadium (chaotic), quadrupole (mixed) shapes.
In the circular dielectric cavity the SPTD has an algebraic long time behavior, $\sim t^{-2}$
in both TM and TE cases, but shows very different short time behavior due to the existence 
of the Brewster angle in TE case where exponential short time behavior is shown. 
The SPTD for a stadium-shaped cavity decays exponentially, and
the exponent has a close relation to the steady probability distribution (SPD). 
In the large $n$ limit, the SPD can be approximated by an assumption of a partial ergodicity, 
a uniform distribution over a specific part of phase space, which gives a correct description of
the exponent in both TM and TE cases. 
We have also discussed about the SPTD for the quadrupolar deformed cavity and
shown that the long time behavior can be algebraic or exponential, depending on the location of islands.

\section*{Acknowledgments}

This work is supported by Creative Research Initiatives of the Korean Ministry
of Science and Technology.

\section*{Appendix}

Here we present the analytical expression of the degree of openness (see Eq.(\ref{gammaT}))
for TM wave.
The degree of openness is defined by
\begin{eqnarray}
I \equiv \int_{-1/n}^{1/n} dp \,\,T(p),
\end{eqnarray}
where $T(p)=1-R(p)$, $R(p)$ is given in Eq.(\ref{Rtheta}). This integral can be
expressed by an analytical function as 
\begin{eqnarray}
I &=& {{4} \over {(n^2 - 1)^2}}
[B({{1} \over {2}},{{3} \over {2}})F(-{{3} \over {2}},{{1} \over {2}};2;{{1} \over {n^2}}) n^2 \nonumber\\
&&+B({{1} \over {2}},{{5} \over {2}})F(-{{1} \over {2}},{{1} \over {2}};3;{{1} \over {n^2}})  
-{{40} \over {15}}n +{{8} \over {15}}{{1} \over {n}}] \nonumber \\
&\simeq& 2 \pi n^{-2},
\end{eqnarray}
where $B(x,y)$ is the beta function and  $F(\alpha,\beta;\gamma;z)$ 
the Gauss hypergeometric function \cite{Gra00}.

For TE wave, only difference is the replacement of $R(p)$ by  $R_{TE}(p)$ 
of Eq.(\ref{RTE}), and the result is
\begin{equation}
I\simeq 4\pi n^{-2}
\end{equation}
for the large $n$ limit based on a numerical calculation.

\end{document}